\documentclass[preprint,5p,sort&compress]{elsarticle}  % model 5p journal style

\usepackage{graphicx}
\usepackage{wrapfig}
\usepackage{multirow}
\usepackage{epstopdf}
\usepackage{url}
\usepackage{slashed}
\usepackage{caption}
\usepackage{subcaption}
\usepackage[utf8]{inputenc}  %permits extended character set

\begin{document} 

\newcommand{\ubarTwo}{\overline{u}_2}
\newcommand{\ubarFour}{\overline{u}_4}
\newcommand{\gE}{g_e}
\newcommand{\epOneSlash}{\slashed{\epsilon}_1}
\newcommand{\gmuUp}{\gamma^{\mu}}
\newcommand{\gmuDown}{\gamma_{\mu}}
\newcommand{\pZeroSlash}{\slashed{p}_0}
\newcommand{\pOneSlash}{\slashed{p}_1}
\newcommand{\pTwoSlash}{\slashed{p}_2}
\newcommand{\pThreeSlash}{\slashed{p}_3}
\newcommand{\pFourSlash}{\slashed{p}_4}
\newcommand{\eGamma}{E_{\gamma}}
\newcommand{\rxnThree}{$\gamma p \rightarrow p \pi^- X$}
\begin{frontmatter}
\title{Design and construction of a high-energy photon polarimeter}

\author{M.~Dugger, B.~G.~Ritchie, N.~Sparks,
K.~Moriya, R.~J.~Tucker, R.~J.~Lee, B.~N.~Thorpe, T.~Hodges}
\address{Arizona State University, Tempe, AZ  85287-1504}

\author{F.~J.~Barbosa, N.~Sandoval}
\address{Thomas Jefferson National Accelerator Facility, Newport News, VA  23606}

\author{R.~T.~Jones}
\address{University of Connecticut, Storrs, Connecticut 06269}
\date{\today}

\begin{abstract}
We report on the design and construction of a high-energy photon
polarimeter for measuring the degree of polarization of a linearly-polarized photon beam. 
The photon 
polarimeter uses the process of pair production on an atomic
electron (triplet production). The azimuthal distribution of
scattered atomic electrons following triplet production yields information regarding the degree
of linear polarization of the incident photon beam.
The polarimeter, operated in conjunction with a pair spectrometer,
uses a silicon strip detector to measure the recoil electron
distribution resulting from triplet photoproduction in a beryllium
target foil. The analyzing power $\Sigma_A$ for the device using a
75 $\rm{\mu m}$ beryllium converter foil is about 0.2, 
with a relative systematic uncertainty in $\Sigma_A$ of 1.5\%.
\end{abstract}


\end{frontmatter}

\section{Introduction} \label{sec:intro}
Multi-GeV polarized photon beams provide a precision tool 
for probing excitations of the quark and gluon substructures of mesons and baryons.  
A common approach for producing multi-GeV linearly-polarized photon beams
is the coherent bremsstrahlung process, in which a high-energy electron
beam undergoes bremsstrahlung within an oriented diamond crystal.
Enhancements in the resulting photon spectrum possess linear polarization. 
The degree and direction of linear polarization are directly controlled by 
the relative orientation of the diamond symmetry axes with respect to the incident electron
beam.

We report here the development of a photon beam polarimeter
based on measuring the polarization of a multi-GeV photon beam
using the so-called ``triplet photoproduction'' process \cite{Mork1967,maxScreen}. 
Such a ``triplet polarimeter" determines the degree of polarization
of the incident photon beam by using the process of triplet photoproduction.
In the triplet photoproduction process, the polarized photon beam
interacts with the electric field of an atomic electron
(rather than the field of the atomic nucleus)
within the material of a production target,
and produces a high energy electron-positron pair
through the pair production process.
The atomic electron on which the pair production took place
then recoils with sufficient momentum to leave the atom and,
if the recoil momentum is great enough and the production target
thin enough,
the electron can leave the target material altogether.
Any momentum of the electron-positron pair transverse to the incoming
photon beam must be compensated by the momentum of the recoil electron.
Typically the  momentum of the recoil electron is much smaller than that of either of the
pair-produced leptons,
so the recoil electron can attain a large polar scattering angle relative
to the axis determined by the incoming photon beam.
When coupled with the trajectory and energy information of the
lepton pair, the azimuthal angular distribution of the recoil electron 
can provide a measure of the photon beam polarization.

This report is organized in the following fashion.
We initially provide a description of the triplet photoproduction
process in terms of quantum electrodynamics (QED),
and show that such a description accurately represents
the experimentally observed cross section for the process.
Next, the design of the triplet polarimeter (TPOL) 
in use within Hall D of the Thomas Jefferson National Accelerator Facility (Jefferson Lab)
is discussed in order
to indicate how considerations of the features of Hall D are reflected
in the design of the device, 
followed by a discussion of details of the construction of the TPOL.  
We then describe simulations of 
the TPOL performance in order to estimate the accuracy to which the photon beam polarization can be
determined with the device.

\section{The triplet photoproduction process} \label{sec:QED}
The cross section for triplet photoproduction can be written as
 $\sigma_t = \sigma_0\left[1 - P \Sigma \cos(2 \phi)\right]$ for a polarized photon beam,
where $\sigma_0$ is the unpolarized cross section, $P$  the photon beam
polarization,
$\Sigma$ the beam asymmetry for the process, and $\phi$ the azimuthal angle of the
trajectory of the recoil electron with respect to 
the plane of polarization for the incident photon beam.
To determine the photon beam polarization, the azimuthal distribution
of the recoil electrons is recorded and fit to the function
$A\left[1 - B \cos(2 \phi)\right]$, where the variables $A$ and $B$ are
parameters of the fit. 
In principle, once $B$ has been extracted from the data,
the degree of photon beam polarization is given by $P = B/\Sigma$.
The triplet photoproduction process is governed solely by QED,
so the beam asymmetry $\Sigma$ can be directly calculated to
leading order in $\alpha_{QED}$.
We now turn to a description of how such a calculation has been performed
for this report.

\subsection{QED diagrams in the triplet photoproduction process}

\begin{figure}[h]
\centering
\includegraphics[width=0.48\textwidth]{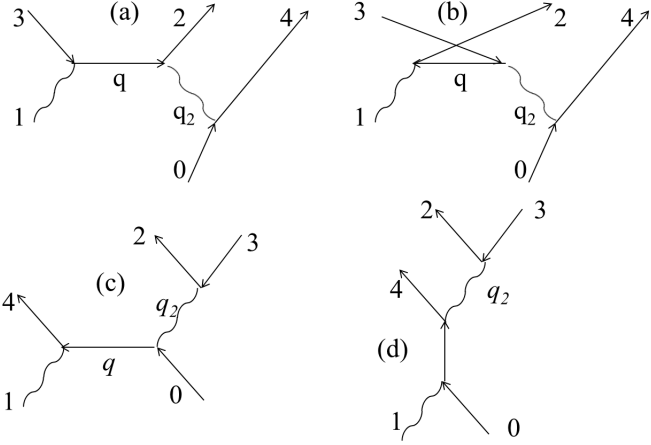}
\caption{The diagrams involved in a QED calculation of the triplet photoproduction process.
This figure illustrates one-half
of the Feynman diagrams involved in the triplet photoproduction process;
the remaining half is obtained by exchanging the electron at 2 with the electron at 4.
Diagrams (a) and (b)
are referred to as ``$\gamma \gamma$-like", while diagrams (c) and (d) are
said to be  ``Compton-like."
For the figures, line 0 represents the target electron,
line 1  is the incident photon,
lines 2 and 3 are the  pair-produced electron and positron, respectively,
and line 4 is the recoil atomic electron.
\label{fig:fdia}}
\end{figure}

A QED calculation of the triplet photoproduction process includes all
8 tree-level QED
diagrams shown in Fig.~\ref{fig:fdia}, 
with corrections due coherent scattering
(also referred to as the screening correction)
included. We now discuss the various terms in Fig.~\ref{fig:fdia} in turn,
indicating the contribution of each to the full calculation. 

\subsubsection{$\gamma \gamma$-like diagrams}
The diagrams (a) and (b) in Fig.\ \ref{fig:fdia} are referred to here as
``$\gamma \gamma$-like." These two diagrams look like the reaction
$\gamma \gamma \rightarrow e^+ e^-$, where one
of the $\gamma$ legs is connected to an electron at position 0 that scatters to position 4.

The matrix elements for the $\gamma \gamma$-like diagrams are given by

\begin{eqnarray}
& -i M_a \equiv (i \gE)^3 \left( \frac{-i}{q_2^2} \right) \times \nonumber \\
& \left[ \ubarTwo \gmuUp i
\left( \frac{\pOneSlash - \pThreeSlash + m}{(p_1 - p_3)^2 - m^2} \right) \epOneSlash v_3 \right]
\left[ \ubarFour \gmuDown u_0 \right],
\label{eqn:ma}
\end{eqnarray}

\begin{eqnarray}
& -i M_b  \equiv (i \gE)^3 \left( \frac{-i}{q_2^2} \right) \times \nonumber \\
&  \left[ \ubarTwo \epOneSlash i
\left( \frac{\pTwoSlash - \pOneSlash + m}{(p_2 - p_1)^2 - m^2} \right) \gmuUp v_3 \right]
\left[ \ubarFour \gmuDown u_0 \right],
\label{eqn:mb}
\end{eqnarray}
where $\gmuDown$ represents the Dirac matrices,
$m$ the electron mass, $q^2_2$ the mass of the virtual photon,
$p$ ($\slashed{p}$) the four-momentum (product of four-momentum with
the Dirac matrices), $u$ and $v$ represent spinors, $\overline{u}$ is an
adjoint spinor, $\epOneSlash$ is the product of
incident photon polarization and
Dirac matrices, and the coupling constant $\gE$ is equal to $\sqrt{4 \pi \alpha}$,
with $\alpha$ being the fine structure constant.
The subscripts 0, 1, 2, 3, and 4 represent the target electron,
incident photon, outgoing electron, outgoing positron, and
recoil electron, respectively.

The matrix elements for the crossed $\gamma \gamma$-like diagrams
(which are not shown in Fig.~\ref{fig:fdia})
are found by switching legs 2 and 4 of diagrams (a) and (b) in Fig.~\ref{fig:fdia}.
Those matrix elements are written as
\begin{eqnarray}
& -i M_{a2} \equiv (i \gE)^3 \left( \frac{i}{q_2^2} \right) \times \nonumber \\
& \left[ \ubarFour \gmuUp i
\left( \frac{\pOneSlash - \pThreeSlash + m}{(p_1 - p_3)^2 - m^2} \right) \epOneSlash v_3 \right]
\left[ \ubarTwo \gmuDown u_0 \right],
\label{eqn:ma2}
\end{eqnarray}

\begin{eqnarray}
& -i M_{b2} \equiv (i \gE)^3 \left( \frac{i}{q_2^2} \right) \times \nonumber \\
& \left[ \ubarFour \epOneSlash i
\left( \frac{\pFourSlash - \pOneSlash + m}{(p_4 - p_1)^2 - m^2} \right) \gmuUp v_3 \right]
\left[ \ubarTwo \gmuDown u_0 \right].
\label{eqn:mb2}
\end{eqnarray}

\subsubsection{Compton-like diagrams}
Diagrams (c) and (d) in Fig.\ \ref{fig:fdia} are referred to here as ``Compton-like."
These two diagrams look like the reaction $\gamma e \rightarrow \gamma e$, where the
scattered $\gamma$ leg is connected to an electron-positron creation vertex.

The matrix elements for the Compton-like diagrams are given by

\begin{eqnarray}
& -i M_c \equiv (i \gE)^3 \left( \frac{-i}{q_2^2} \right) \times \nonumber \\
& \left[ \ubarFour \epOneSlash i
\frac{\pFourSlash - \pOneSlash + m}{(p_4 - p_1)^2 - m^2} \gmuUp u_0 \right]
\left[ \ubarTwo \gmuDown v_3 \right],
\label{eqn:mc}
\end{eqnarray}

\begin{eqnarray}
& -i M_d \equiv (i \gE)^3 \left( \frac{-i}{q_2^2} \right) \times \nonumber \\
& \left[ \ubarFour \gmuUp i
\frac{\pOneSlash + \pZeroSlash + m}{(p_1 + p_0)^2 - m^2} \epOneSlash u_0\right]
\left[ \ubarTwo \gmuDown v_3 \right].
\label{eqn:md}
\end{eqnarray}

Switching legs 2 and 4 in diagrams (c) and (d) in Fig.\ \ref{fig:fdia}
gives us the crossed Compton-like diagrams:
\begin{eqnarray}
& -i M_{c2} \equiv (i \gE)^3 \left( \frac{i}{q_2^2} \right) \times \nonumber \\
& \left[ \ubarTwo \epOneSlash i
\frac{\pTwoSlash - \pOneSlash + m}{(p_2 - p_1)^2 - m^2} \gmuUp u_0 \right]
\left[ \ubarFour \gmuDown v_3 \right],
\label{eqn:mc2}
\end{eqnarray}

\begin{eqnarray}
& -i M_{d2} \equiv (i \gE)^3 \left( \frac{i}{q_2^2} \right) \times \nonumber \\
& \left[ \ubarTwo \gmuUp i
\frac{\pOneSlash + \pZeroSlash + m}{(p_1 + p_0)^2 - m^2} \epOneSlash u_0\right]
\left[ \ubarFour \gmuDown v_3 \right].
\label{eqn:md2}
\end{eqnarray}

\subsubsection{Total matrix element for triplet photoproduction}
By including the matrix elements shown in Fig.~\ref{fig:fdia} and
those given by exchanging lines 2 and 4,
the full matrix element for the triplet photoproduction is calculable.
The total matrix element $M_{\rm{tot}}$ is simply the sum of
the eight matrix elements provided in Equations~\ref{eqn:ma}-\ref{eqn:md2} above:
\begin{eqnarray}
M_{\rm{tot}} & = & M_a + M_b + M_c + M_d + \nonumber \\
& & M_{a2} + M_{b2} + M_{c2} + M_{d2}.
\label{eqn:mTot}
\end{eqnarray}

\subsubsection{Screening correction}
\label{section:sc}
The atomic electron on which the triplet photoproduction process
takes place is embedded in a target material of atomic nuclei and
other electrons; in this document, this material is referred to as the converter.  
The effects of the presence of these additional particles on the 
triplet production process are embodied in 
a screening function $S(q)$ 
(sometimes called an incoherent scattering function), 
as discussed in Refs.~\cite{maxScreen,HubbellScreen}.
The correction due to the screening function $S(q)$ for triplet
production on hydrogen is related to the
atomic form factor $F(q)$ such that
\begin{eqnarray}
S(q) = 1 - F^2(q) ,
\label{eqn:Screen}
\end{eqnarray}
where
$F(q) = \left( 1 + a^2 q^2 /4  \right)^{-2}$, 
$q$ is the momentum transfer to (i.e., the recoil momentum of) the atomic electron, 
and $a$ is the Bohr radius~\cite{maxScreen}.

The ratio of the triplet production cross section to the pair production
cross section on an atom within the converter scales as the atomic number $Z^{-1}$.
As a practical matter, then,
the specific material chosen should have a low $Z$
in order that this ratio is more favorable to triplet production.  
For the triplet detector described in this document,
a beryllium converter foil was used, 
which has the lowest $Z$ for a non-reactive metal ($Z=4$).
Thus, a beryllium screening function~\cite{HubbellScreen} was used for 
evaluation of the triplet photoproduction cross section below.

\subsubsection{Evaluation of the cross section}

Using the above results,
the cross section can be evaluated by using Eq.~\ref{eqn:mTot}
in the event generator of a simulation for the process.
We proceed in the following fashion.
The phase space for the reaction is randomly generated event-by-event.
%The phase space for the reaction is randomly generated event-by-event using an algorithm provided by
%Richard Jones. 
For each event, the matrix element
$|M_{\rm{tot}}|^2$ is constructed from Eq.~\ref{eqn:mTot}
and then averaged over initial spins and summed over final spins to obtain
$\left<|M_{\rm{tot}}|^2\right>$.
For the case where the incident photon is
polarized, the spin state of the photon is fixed in value while averaging
over the initial spin states of the target electron. 
The cross section is obtained using
\begin{eqnarray}
\frac{d\sigma}{d\Omega} =
\frac{(2 \pi)^2}{4 k m} \left<|M_{\rm{tot}}|^2\right> \rho_f S(q),
\label{eqn:total}
\end{eqnarray}
where $k$ is the incident photon energy and $\rho_f$ represents
the final-state phase space. Each event is then assigned
a weight equal to the calculated cross section.

\begin{figure}[t]
\centering
\includegraphics[width=\linewidth]{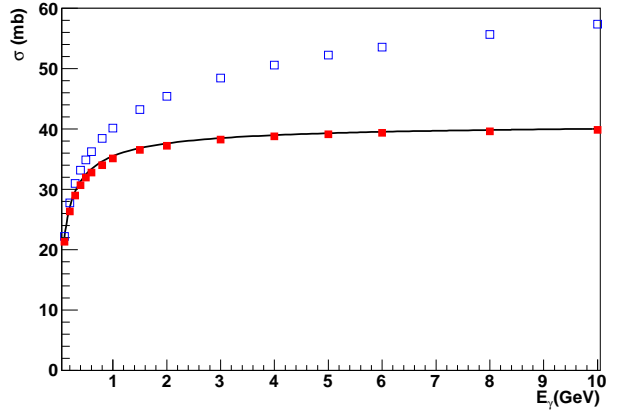}
\caption{{\small{
The total cross section for triplet photoproduction for beryllium
as a function of incident photon energy. 
The red filled (blue open) squares
represent generator results with (without) screening factor $S(q)$.
The solid black line represents the NIST values \cite{NISTxcom}.
%represents the value of the total cross section
%given by NIST.
}}}
{\label{fig:NISTComp}}
\end{figure}

The calculated total cross section as a function of incident photon energy using Eq.~\ref{eqn:total}  
with and without consideration of the screening correction $S(q)$
is shown in  Fig.~\ref{fig:NISTComp}. 

\subsubsection{Comparison with experimental results}
The results from Eq.~\ref{eqn:total} are compared
to the National Institute of Standards and Technology (NIST)
cross sections for beryllium \cite{NISTxcom} in Fig.~\ref{fig:NISTComp}.
In that figure, the solid black line represents the value of the total cross section given by NIST,
and the results from the event generator are shown as squares;
the results without the  screening
function $S(q)$ are shown as blue open squares,
while red filled squares denote the results where that
 screening function has been included.
As seen in the figure, the agreement between the event generator and NIST is excellent
when the  screening function is included.
The agreement seen in Fig.~\ref{fig:NISTComp} indicates that the 
triplet photoproduction process has been
properly described by Eqs.~\ref{eqn:ma}-\ref{eqn:Screen} above.

\section{Design of a triplet polarimeter for Hall D} \label{sec:design}
We describe in this section the design strategy for the triplet polarimeter TPOL, 
which detects the recoil electrons from the triplet process in order
to measure the beam polarization of an intense 
($10^7-10^8$ photons/s in the coherent peak) 
linearly-polarized photon beam
with photon energies up to 9 GeV.
The GlueX experiment~\cite{Ghoul:2015ifw}
uses the linearly-polarized photon beam 
available in  the newly-constructed Hall D at the Thomas Jefferson National Accelerator Facility (Jefferson Lab). 
GlueX will explore the properties of hybrid mesons, in which the gluonic field contributes directly
to the quantum numbers of the mesons~\cite{Meyer:2015eta}.
These explorations demand precise knowledge
of the incident photon beam polarization.

\begin{figure}[t]
\centering
\includegraphics[width=0.48\textwidth]{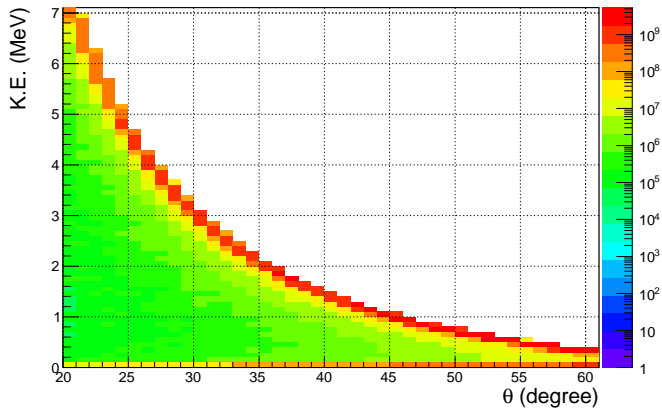}
\caption{
Cross-section-weighted kinetic energy versus polar angle for the recoil electron.
\label{fig:keTheta}}
\end{figure}

The central component in a triplet polarimeter is the charged particle detector
used for intercepting the recoil electrons from the photoproduction process. 
When considering the geometry of potential detectors for the recoil electron, 
the expected angular distribution and kinetic
energies of the those electrons must be investigated. 
In Fig.~\ref{fig:keTheta}, the
cross-section-weighted kinetic energy versus polar angle $\theta$ for the
recoil electron is plotted for incident photons with
energies between 8 and 9 GeV. As seen in the figure, the
polar angle increases as the recoil electron kinetic energy decreases.
Thus, simply by inspection of Fig.~\ref{fig:keTheta}
and considering the minimum kinetic energy for a recoil electron 
emerging from the production target, 
the maximum polar angle for a possible detector system can be
estimated. 
For example, if the
minimum desired kinetic energy for the recoil electron is 1 MeV,
then the maximum polar angle of the detector should be about $\theta = 45^\circ$.

With this initial design parameter established, the general dimensions of the
detector geometry then can be investigated by simulation,
using the same QED-based approach outlined in the previous section
to generate events arising from the triplet photoproduction process. 
For purposes of simulation, events were generated with 
an idealized incident photon beam, 
where photons were polarized 100\% in the $x$-direction with beam energies between 
8 and 9 GeV. 
With this ideal beam, a histogram with 36 bins in azimuthal angle was created 
and filled with cross-section-weighted events. 

\begin{figure}[h]
\centering
\includegraphics[width=0.48\textwidth]{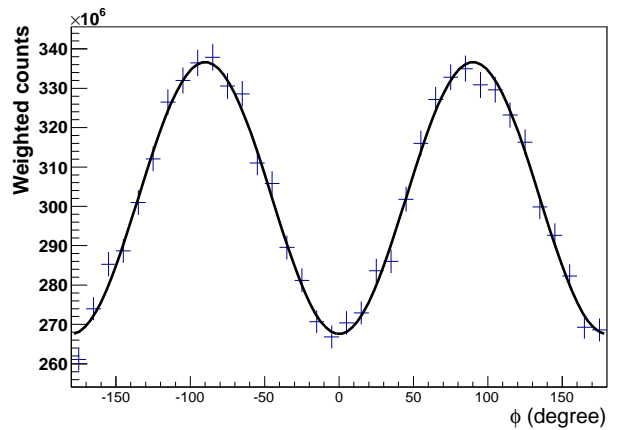}
\caption{
Cross-section-weighted simulated triplet photoproduction events versus azimuthal angle $\phi$
for recoil electrons with polar angle $\theta = 30^\circ$. 
Also shown by the solid curve is a fit to the azimuthal distribution 
using $A[1 - \Sigma \cos(2\phi)]$, with $A$ and $\Sigma$ as fit parameters.
\label{fig:sigPlot}}
\end{figure}

\begin{figure}[h!]
\centering
\includegraphics[width=0.48\textwidth]{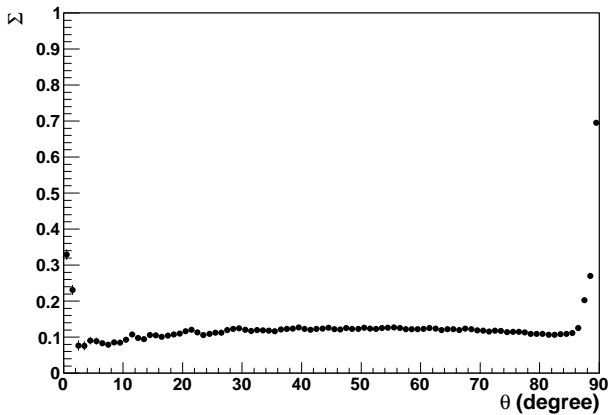}
\caption{
Simulated beam asymmetry $\Sigma$ as a function of polar angle,
with no restrictions on the kinetic energy of the recoil electron. 
\label{fig:anaTheta}}
\end{figure}

Fig.~\ref{fig:sigPlot} 
shows the cross-section-weighted counts versus azimuthal angle $\phi$ for  
recoil electrons with polar angle $\theta = 30^\circ$. 
Since the polarization is set to $P=1$ in these simulations,
the resulting azimuthal distribution was fit 
to the functional form $A[1 - \Sigma \cos(2\phi)]$, 
where $A$ and $\Sigma$ were the fit parameters.
Shown in Fig.~\ref{fig:anaTheta} is the beam asymmetry (parameter $\Sigma$ 
from the fit) as a function of polar angle $\theta$. 
The asymmetry is largest when 
the recoil electron has polar angles near $0^\circ$ and near $90^\circ$, 
but at large scattering angles the kinetic energy is very small, 
and at small angles one confronts the difficulty of placing 
a charged particle detector within a couple degrees of 
the beamline without moving that detector far away from the reaction vertex
(and thereby greatly increasing the size of the detection system).
However, Fig.~\ref{fig:anaTheta} indicates a relatively flat behavior at most
intermediate polar angles, which in turn indicates a relative insensitivity to small
deviations in the recoil electron polar angle when determining the beam asymmetry.
When combined with the initial choice of the maximum polar angle from Fig.~\ref{fig:keTheta},
the behavior seen in Fig.~\ref{fig:anaTheta} provides a reasonable method for
determining the minimum and maximum polar angles
for the recoil electron detector, and, thus, the dimensions of that detector. 

The beam asymmetry for  triplet photoproduction  
also depends on the kinematics of the produced $e^+$ $e^-$ pair. 
Thus, a spectrometer is needed to determine the energies and trajectories 
for the produced electron and positron
in order to fully specify the kinematics for  the reaction.
In Hall D, such a ``pair spectrometer"~\cite{Barbosa:2015bga} 
is located within the Hall D experimental area,
and the converter is placed approximately 7.5 m upstream of the focal plane of that device. 
The event generator can be used to explore the
dependence of $\Sigma$ on the energy difference $\Delta E$ between the kinetic energies of
each member of the produced $e^+$ $e^-$ pair.
As was seen in Fig.~\ref{fig:keTheta}, the kinetic energy of the recoil electron is on the order
of a few MeV for polar angles above 20$^\circ$. 
This means that, for large recoil polar angles,
nearly all the multi-GeV incident photon energy 
is divided between the members of the produced $e^+$ $e^-$ pair.

\begin{figure}[h]
\centering
\includegraphics[width=0.48\textwidth]{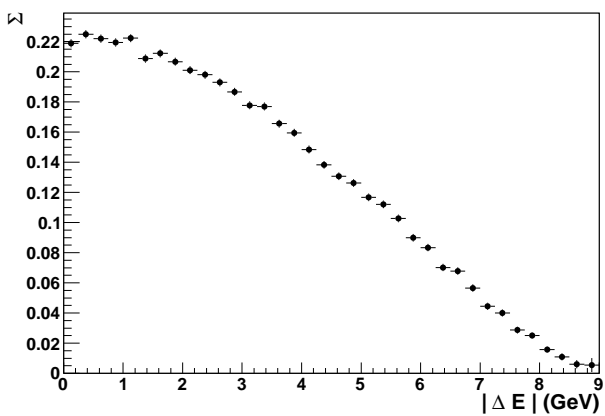}
\caption{
Dependence of the beam asymmetry $\Sigma$ on $|\Delta E|$, 
the difference between the kinetic energies of the produced $e^+$ and $e^-$.
\label{fig:anaDelta}}
\end{figure}

The simulated dependence of the beam asymmetry on the absolute value
$| \Delta E | = | E_+ - E_- |$, where $E_+$ is the energy of the positron and $E_-$ is
the energy of the produced electron, is shown in Fig.~\ref{fig:anaDelta}.
As the energies of the produced leptons become more equal (i.e., as $|\Delta E|$ becomes closer to 0),
the beam asymmetry increases to a maximum of about 0.22. 
If a cut is applied such that events above a particular value of $|\Delta E|$ are removed, however, 
there necessarily will be a corresponding decrease in statistical accuracy in the measurement of 
that asymmetry. 
A particular choice for a $|\Delta E|$ cut must be considered in terms of its impact 
on the expected statistical uncertainty in the measured beam asymmetry.

\begin{figure}[h]
\centering
\includegraphics[width=0.48\textwidth]{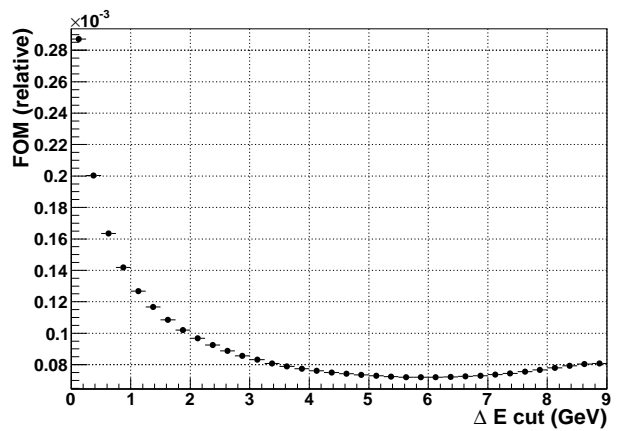}
\caption{
Relative FOM as a function of the cut in energy difference $|\Delta E|$ (i.e., where energy
differences greater than the $|\Delta E|$ value shown on the $x$-axis are neglected in the determination of
relative FOM given on the $y$-axis). For each plot, the generated incident photon energy
was uniformly distributed between 8 and 9 GeV.
\label{fig:fomDelta}}
\end{figure}

The expected statistical uncertainty in the derived photon beam polarization is
\begin{eqnarray}
\frac{\sigma_P}{P} = \frac{1}{\sqrt{N}} \sqrt{\frac{2}{\Sigma^2 P^2} - 1} ,
\end{eqnarray}
where $P$  is the polarization, $\sigma_P$ the standard deviation in that quantity, $N$ 
the number of triplet events surviving cuts, and $\Sigma$ the beam asymmetry.
For small values of $\Sigma P$.
\begin{eqnarray}
\sigma_P  \approx \frac{\sqrt{2}}{\Sigma \sqrt{N}}.
\end{eqnarray}
Thus, if we define a figure of merit (FOM) with the expression $\rm{FOM} = 1 / \sqrt{\Sigma^2 N}$,
then FOM is directly proportional to the expected uncertainty in $\Sigma P$, 
with smaller values of FOM being more desirable. 
To relate the FOM to the derived value of 
$\Sigma$ from the event generator, we create a ``relative figure of merit'' $\rm{FOM_r}$ 
such that $\rm{FOM_r} = 1 / \sqrt{\Sigma^2 N_w}$, where $N_w$ is the 
number of cross-section-weighted events from the event generator. 
Fig.~\ref{fig:fomDelta} shows 
$\rm{FOM_r}$ as a function of cut values in $| \Delta E |$, with the best 
$\rm{FOM_r}$ occurring when a cut is applied at about 6 GeV (events with $| \Delta E |$
greater than 6 GeV being removed from the analysis).

\label{sc:beamOffset}
\begin{figure}[h]
  { \centering
    \includegraphics[width=1.0\linewidth]{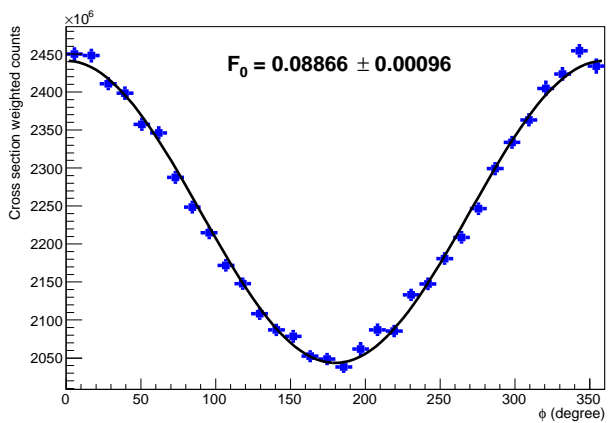}
    \caption{\small{
        The blue points are unpolarized Monte Carlo data
        with the incident photon-beam offset from the detector center by
        1.0 mm in the x-direction.
        The solid line is a fit of the data with the function $A[1+F_0\cos(\phi)]$,
        where $A$ and $F_0$ are parameters of the fit. The value of $F_0$ is shown
        on the plot and has a value of 0.089(1).
    }}
    \label{fig:f0}}
\end{figure}
\begin{figure}[h]
 { \centering
  \includegraphics[width=1.0\linewidth]{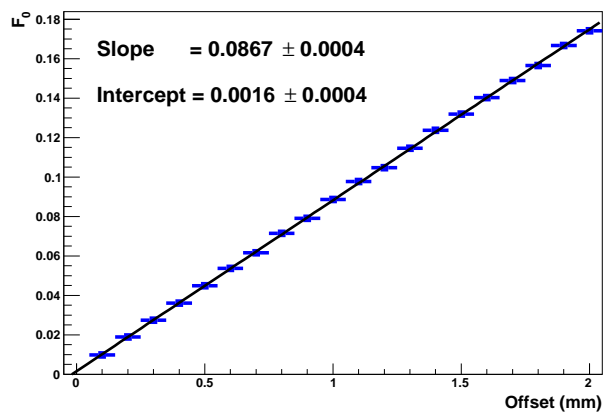}
  \caption{\small{
                  Plot of the fit parameter $F_0$ in \ref{fig:f0} versus  beam offset in mm. The blue points are from fit results,
                  and the solid line is a first degree polynomial fit to the data, where
                  the slope and intercept are displayed on the figure.
      }}
  \label{fig:f0VsO}}
\end{figure}

As a final design consideration, it is important to note that the observed recoil electron asymmetry 
will be altered if the recoil electron detector and the incident photon beam are not
precisely coaxial.
In practice, in our case, actual beam offsets should be
rather small (typically less than a few hundred $\mu$m at Jefferson Lab) 
due to the care taken both in aligning the detector with the beamline
and in delivering the incoming electron beam to the diamond radiator.
Nonetheless, the effect of an offset in the photon beam spot position 
transverse to the recoil electron detector center
must be explored. We simulated the detector response for different
values of photon beam offset using unpolarized generated events.
For that purpose, an unpolarized photon beam spot uniformly 
distributed over a 5 mm disk was simulated.
Twenty evenly-spaced beam offsets ranging from 0.1 mm to 2.0 mm were analyzed.
Fig.~\ref{fig:f0} shows a plot of cross-section-weighted events in blue
for events generated with a 1-mm beam offset in the x-direction. Additionally, the
figure shows a solid black line that represents a fit to the data, where the fit function
is given by $A[1+F_0\cos(\phi)]$, with $A$ and $F_0$ being fit parameters.
As seen from the figure, the function fits the result well, and
in this case, produces a value of $F_0 = 0.089(1)$.
In a similar manner, values of $F_0$ were determined for each of the generated beam offsets.
The dependence of $F_0$ on the amount of generated beam offset was found to be 
remarkably linear,
as seen in Fig.~\ref{fig:f0VsO}.
The beam offset mainly causes the observed azimuthal distribution to exhibit 
first-order
Fourier moments, and thus, due to orthogonality, the beam offset has only a small
impact on the extraction of the second cosine moment used
to determine the beam asymmetry.

\section{TPOL construction} \label{sec:construction}
Based on the design considerations outlined in the previous section,
we have constructed a system for detection of recoil electrons following
triplet photoproduction. 
The various elements of TPOL and details of construction are discussed in this section. 
The system consists of (1) a converter tray and positioning assembly, 
which holds and positions a beryllium foil converter within which the
triplet photoproduction takes place; 
(2) a silicon strip detector (SSD) to detect the recoil electron from triplet photoproduction,
providing energy and azimuthal angle information for that particle; 
(3) a vacuum housing containing the production target and SSD,
providing a vacuum environment minimizing multiple Coulomb scattering between
target and SSD; and 
(4) the preamplifier and signal filtering electronics within
a Faraday-cage housing.

\subsection{SSD mount, converter tray positioning system, and detector plate}
\begin{figure}[t]
\centering
\includegraphics[width=0.5\textwidth]{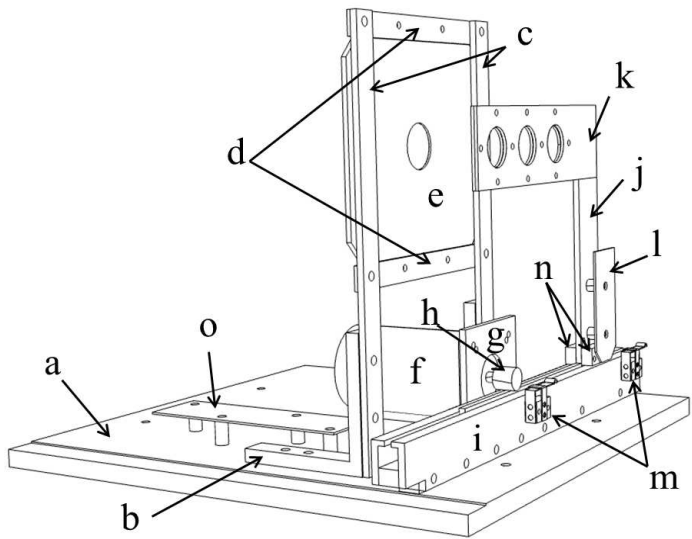}
\caption{
Drawing of removable detector plate assembly (fasteners and wires not shown). 
The components are shown in the orientation used for assembly and bench testing
of the SSD and converter positioning system.
During normal operation, the system is inverted for insertion into the vacuum chamber. 
\label{fig:detPlate}}
\end{figure}

As shown in Fig.~\ref{fig:detPlate},
several components of TPOL are mounted on 
a 10 $\times$ 10 in$^2$ removable detector plate
(labeled ``a" in the figure), fabricated from 0.375-in-thick aluminum.
The detector plate permits those components to be mounted, serviced, and tested together outside
the vacuum chamber (described below) as needed, 
and then to be installed within the vacuum chamber for normal operation. 
The figure shows the detector plate in the orientation used for servicing and 
bench testing;
during normal operation, the plate (along with the detector and converter tray assemblies)
is inverted and slid onto mounting rails within the vacuum chamber,
as discussed below.

The SSD (described below) is mounted on a rigid mounting frame,
as seen in Fig.~\ref{fig:detPlate}.
Firmly attached to the detector plate (``a"),
the SSD mounting frame consists of two leg support brackets (``b") 
to which are attached a pair of detector frame legs (``c").
A pair of detector cross bars (``d") are attached at right angles to 
the detector frame legs,
%those leg pieces,
and the SSD printed circuit mounting board (``e") is affixed to 
the detector cross bars. 
All pieces of the mounting frame are made of aluminum. 
  
The polarimeter is installed within the collimator cave
in the Hall D photon beam line during normal operation,
which is a high radiation zone when the photon beam is incident on the device.
An ability to remotely place and position converters of various thicknesses into the incident photon beam
(or to remove them altogether) while maintaining the detector system under vacuum
is desirable.
This capability is provided by a converter positioning system shown in Fig.~\ref{fig:detPlate}
(items ``f" through ``n").  
A positioning stepper motor (``f"; described more fully below) is attached to the detector plate (``a") 
by a mounting bracket (``g") fixed to the plate. 
The positioning motor has a metric 0.5 module gear (``h"; gear teeth are not represented
in the figure) that adjusts the position of a 0.5 module metal rail assembly contained
within a rail guide (``i"). 
The rail assembly consists of the metal rail itself, sandwiched between two 
0.125-in-thick aluminum pieces filed such that the rail assembly 
moves smoothly inside the rail guide.
The converter tray (``j") is a single, L-shaped piece
of aluminum, with one leg attached to the rail
and the other leg containing holes for up to three different converter foils. 
A removable cover (``k") allows the converter foils to be
installed and held in place on the converter tray.

The converter tray is moved by the stepper motor so as to 
position the converter of choice in the photon beam,
with the motor operated remotely
by the motor controller electronics. 
The stepper motor (Phytron VSS 43.200.1.2-UHVG) is
a two-phase stepper motor with 200 steps per revolution.
The motor selected for this purpose can withstand a radiation dose of 1 MJ/kg, %10$^{6}$ J/kg, 
and can operate at pressures as low as 7.5 nTorr. %\times 10^{-12}$Torr.
To ensure that the positioning system does
not move the converter tray beyond acceptable bounds, 
a strike plate (``l") attached to the converter tray through nylon spacers 
will engage one of two limit switches (``m") if the converter tray
moves beyond the preset range of motion. 
These normally-closed vacuum-rated limit switches (allectra 363-SWITCH-01) can operate with currents up to 1 A
and provide, when engaged, an indication to the motor controller.
Two guide blocks (``n") fixed to the converter tray limit unwanted motion
about the axis defined by the rail.
While no fasteners or wires are shown in the figure, 
wires from the limit switches and from the motor are guided underneath
a wire guide plate (``o") attached to the detector plate using steel spacers;
the wire guide plate keeps wires in place when
the detector plate is inverted and placed within the vacuum housing.  

\subsection{Silicon strip detector}
The recoil electrons from triplet photoproduction are detected by 
an S3 double-sided silicon strip detector manufactured by Micron Semiconductor.
The detector provides energy and trajectory information
for those particles. 
The S3 has 32 azimuthal sectors on the ohmic side and
24 concentric rings on the junction side. 
The detector has an outer active diameter of
70 mm and an inner active diameter of 22 mm. 
The thickness of the silicon is 1034 $\mu$m, % microns,
and the material is fully depleted with a bias potential of 165 V.
During normal operation, the SSD is operated 
at the manufacturer's suggested operating voltage of 200 V.

\subsection{Vacuum chamber}
To minimize energy loss and multiple Coulomb scattering of the recoil electrons
as they travel between the converter foil and the SSD,
the detector plate and all components shown in Fig.~\ref{fig:detPlate} 
are placed within a vacuum chamber manufactured by the Kurt J. Lesker 
Company.
(As noted above, the entire assembly shown in Fig.~\ref{fig:detPlate}
is inverted before installing in the vacuum chamber.)
The vacuum chamber, with interior volume 12 $\times$ 12 $\times$ 12 in$^3$,
is physically located in the Hall D
collimator cave just upstream of the pair spectrometer.
Using the manufacturer's web application, the 
design of one of their standard steel box chambers was modified 
to include ports and flanges for the purposes described below. 
Support brackets for positioning the detector plate and hardware shown in Fig.~\ref{fig:detPlate}
were welded to the interior upper horizontal surface of the chamber by the manufacturer.
The vacuum chamber walls themselves are made of 0.5-in-thick 304 stainless steel.

\begin{figure}[t]
	\begin{minipage}[t]{\linewidth}
  		\centering \includegraphics[scale=0.4]{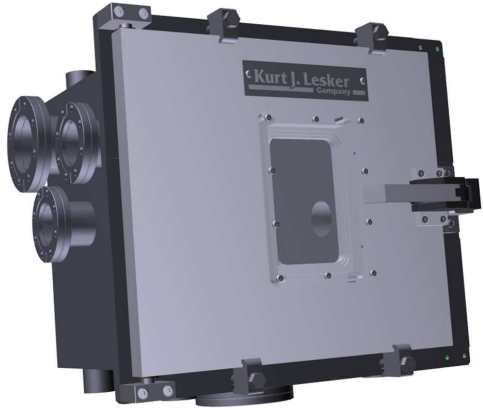}
  		\subcaption{Front view of vacuum chamber}
  		\label{fig:chamfront}
   	\end{minipage}
   	\begin{minipage}[t]{\linewidth}
   		\centering \includegraphics[scale=0.4]{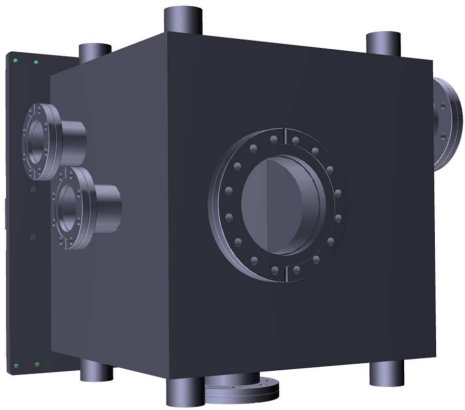}
   		\subcaption{Rear view of vacuum chamber}
   		\label{fig:chamrear}
	\end{minipage}
	\caption{Vacuum chamber housing TPOL components. 
	In (a), the photon beam enters from the right, while in (b) 
	the photon beam enters from the left. The functions of the various ports
	and flanges are discussed in the text. }
	\label{fig:chamber}
\end{figure}

The exterior of the vacuum chamber is illustrated by the drawings provided in Fig.~\ref{fig:chamber}. 
On the downstream side of the chamber (seen at left in Fig.~\ref{fig:chamfront})
are three vacuum conflat flanges. 
The uppermost 4.5-in-diameter flange is used 
for routing cables related to the converter tray positioning 
system (items ``f" through ``n" in Fig.~\ref{fig:detPlate}) described above. 
The lower flange (rotatable, 3.375-in-diameter) provides connection to
the beam pipe, and the remaining flange (3.375 in diameter) is used to connect an 
auxiliary vacuum pump. 
An access door shown in Fig.~\ref{fig:chamfront},
constructed from a 1-in-thick 6061-T6 aluminum plate, 
includes a viewport (3.25$\times$5.25 in$^2$) with a borosilicate glass cover.
The back of the chamber seen in Fig.~\ref{fig:chamrear} has a 6-in-diameter flange   
for routing signal cables from the SSD. 
A flange located on the bottom of the chamber (4.5 in diameter) is 
used for mounting a turbomolecular vacuum pump. 
Two 3.375-in-diameter flanges are located on the upstream side (left side of Fig.~\ref{fig:chamrear})
of the vacuum chamber; the uppermost flange
is for an auxiliary vacuum gauge, while  the other (rotatable) flange is 
used as the port for the photon beam entrance,
and is coupled to the photon beamline. 

\begin{figure}[t]
\centering
\includegraphics[width=0.45\textwidth]{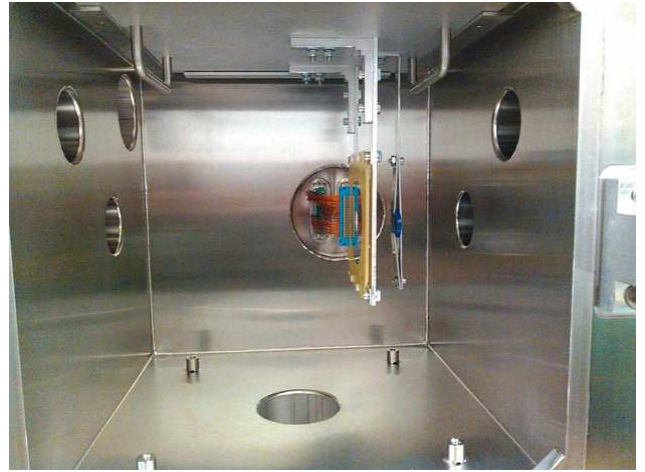}
\caption{(Color online)
Interior view of vacuum chamber, with detachable plate and silicon strip detector
installed.  
(The converter tray positioning system has been removed in this picture.) 
Also visible in the photograph is a 1 $\mu$Ci
$^{137}$Cs radioactive source (blue disk) 
held in place by a temporary source stand used for initial tests.
During normal operation, 
the source and source stand shown are 
removed, and the converter tray positioning system installed. 
\label{fig:insideChamber}}
\end{figure}

An interior view of the vacuum chamber,
with the converter tray positioning system removed, 
is provided in Fig.~\ref{fig:insideChamber}. 
The SSD, attached to a (yellow) printed circuit board, is shown attached to
the SSD mounting frame described above. 
As seen in the figure, handles of 304 tubular stainless steel (Grainger 4LAF8) 
are installed on the detector plate to assist removal and insertion.
Fig.~\ref{fig:insideChamber} also shows Kapton-insulated wires 
(30 AWG, Accu-Glass Products 112739) running
from the far side of the SSD printed circuit board to a 6-in-diameter electrical vacuum feedthrough  flange.
The feedthrough flange contains two 50-pin connectors (Accu-Glass Products 50D2-600). 
The SSD-facing ends of each Kapton wire are mated to the SSD printed circuit board
using a standard wire-to-board connector (Yamaichi NFS64A0111), 
while the ends facing the feedthrough flange 
are crimped to female, gold-plated sockets (Accu-Glass Products 111652),
which have then been inserted into a commercially-available 
50-socket glass-filled dyiathilate high-vacuum D-connector
(Accu-Glass Products 110007). 
One of the D-sub connectors routes signals for the SSD azimuthal sectors, 
while the other D-sub passes the signal lines
for the concentric rings.

\subsection{Detector electronics}
As noted above, signals from the SSD are passed from
the vacuum chamber through an electrical vacuum feedthrough flange 
containing two 50-pin D-sub connectors. 
The face of the  feedthrough flange that is
at atmospheric pressure is housed in a metal box
enclosure serving as a Faraday cage for the preamplifier electronics.
Signals from the downstream (azimuthal sector) side of the SSD
are fed through the vacuum flange into a charge-sensitive preamplifier manufactured by Swan Research.
(At present, wires from the concentric ring side are connected to ground;
a future upgrade will provide preamplifier electronics for the rings.)

\begin{figure}[t]
\centering
\includegraphics[width=0.43\textwidth]{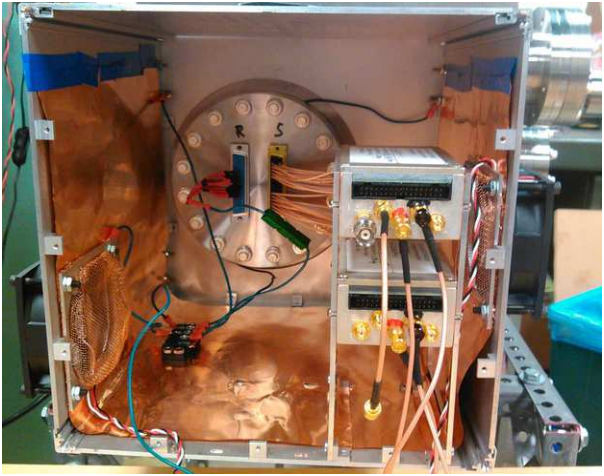}
\caption{(Color online)
Interior view of preamplifier enclosure,
with the front panel removed.
Two sets of preamplifiers are housed in the 
two metal boxes mounted on standoffs
at right in the foreground. 
Thin copper foil is used to line the top, bottom, and side walls of the enclosure.  
\label{fig:preamps}}
\end{figure}

The photograph in Fig.~\ref{fig:preamps} shows an interior view 
of the preamplifier enclosure
attached to the back of the vacuum chamber.
The ring side cables run from the D-sub to ground
through a terminal block (black rectangular box attached to the preamp enclosure
floor), while the sector side cables (RG-178 braided coaxial) run from the D-sub to SMA connectors
on the preamplifiers, which are housed in two metal boxes. 

To reduce exterior electromagnetic signal interference,
the box is lined with a layer of copper foil.
While not easily seen in the figure, the stands that are used to attach the preamp
boxes to the enclosure also are lined with copper foil on the side facing 
the preamp boxes, with that foil making good contact with the preamp boxes and the foil lining
of the enclosure. 
The copper foil is attached to ground at several points 
through wires leading to a terminal block mounted to bottom wall of the enclosure.  
Attached to the left and right outside walls 
of the enclosure are two 80-mm case fans (Enermax UC-8EB)
to provide cooling. 
On the interior side of the holes required for the fans, 
copper mesh (16 strands per inch with 0.011 inch diameter strands)
are installed for additional electromagnetic shielding. 

Each preamplifier box requires a positive and negative 12 volt potential.
In Fig.~\ref{fig:preamps}, SMA cables with end caps marked black are the negative 12 volt
inputs to the preamps. Red markings are above the SMA connectors that
supply the positive 12 volts. The remaining SMA cable on the top 
preamp box is to pass the bias voltage to the sector side of the 
SSD. 
 
In the depletion region of the silicon detector,
an electron hole pair is created for each 3.6 eV of energy deposited.
The electrons are collected on the positive side of the potential
(sector side of the detector), while the holes are swept towards the ring side (ground).

\begin{figure}[h]
\centering
\includegraphics[width=0.3\textwidth]{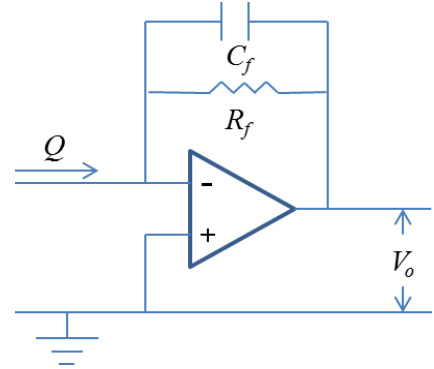}
\caption{
Simplified circuit diagram of the charge-sensitive preamplifier
used to process signals from the silicon strip detector in TPOL.
The feedback capacitance ($C_f$) and resistance ($R_f$)
determine the signal fall time presented at $V_0$. 
\label{fig:gpreamp}}
\end{figure}

The charge-sensitive preamplifiers were manufactured by Swan Research (BOX16CHIDC/CHARGE8V).
A simplified circuit diagram for the charge-sensitive preamplifier 
is given in Fig.~\ref{fig:gpreamp}. 
The feedback capacitance ($C_f$) and resistance ($R_f$)
determine the signal fall time. 
For the TPOL, $C_f$ = 0.2 pF and $R_f$ = 30 M$\Omega$ were chosen,
resulting in a fall time of 6 $\mu$s. 
The rise time for the signal is determined by the charge collection time of the detector 
and by the response time of the preamplifier.
Optimal sensitivity $S_o$ of the preamplifier 
is determined by the feedback capacitance such that $S_o = 1/C_f$, 
which leads to $S_o \approx 250$ mV/MeV. 

Electronic filters and amplifiers were added to the
output of the charge-sensitive preamplifiers in order to reduce unwanted noise
and to provide a small amount of signal shaping before
passing the signals to analog-to-digital converters
within the GlueX electronics.

\subsection{TPOL installation}
\begin{figure}[h]
\centering
\includegraphics[width=0.4\textwidth]{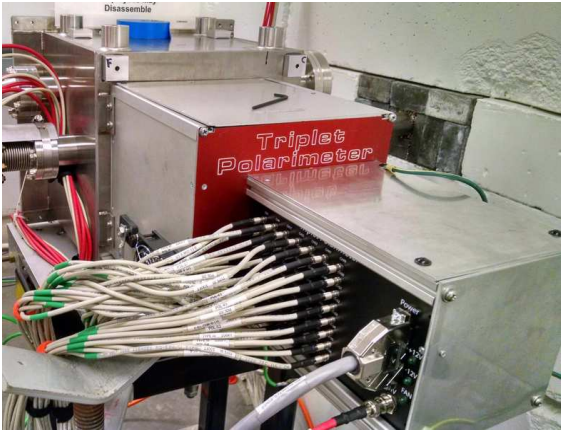}
\caption{(Color online)
Triplet polarimeter installed in the Hall D beamline.
The photon beam transits from left to right in this photograph 
within the beam pipe seen at rear left.
The largest box at rear/left is the TPOL vacuum chamber. 
\label{fig:tripBGR2}}
\end{figure}

A photograph of the assembled TPOL  
installed within the collimator cave in the Hall D beamline is shown
in Fig.~\ref{fig:tripBGR2}, 
appearing as three boxes attached to one another.
The largest box at the rear is the vacuum chamber, while the middle box is 
the preamplifier enclosure.
The smallest box (closest to the viewer) is a signal distribution box that
routes voltages and signals to and from the preamplifiers. 
The cables attached to the 
distribution box are the signal lines (white), high voltage bias (red) and
low voltage (gray). 

In operation,
the TPOL vacuum box is coupled directly to the evacuated beamline
through which the polarized photon beam passes. 
A diamond serves as the coherent bremsstrahlung photon production target 
for the Hall D photon beam, and is located upstream of the TPOL detector.
Upon entering TPOL, the photon beam passes into whichever beryllium converter is
positioned in the beam. 
Triplet photoproduction takes place within the converter material,
with the recoil electron being detected by the SSD within the
TPOL vacuum chamber. Produced electron-positron pairs, 
as well as any photons that did not interact with the converter material,
pass through the downstream port of the vacuum box 
into the evacuated beamline, which in turn
passes through a shielding wall (seen at right in the photograph)
into the Hall D experimental area.
The $e^+ e^-$ pair then enters 
the vacuum box and magnetic field of the GlueX pair
spectrometer, while photons continue
through an evacuated beamline to the target region
of the GlueX detector. 
During normal running conditions, the photon beamline and
TPOL vacuum chamber are operated at approximately 10 $\mu$Torr. 

\section{Analyzing power for TPOL} \label{sect:accuracy}
\subsection{Analyzing power $\Sigma_A$ for a given $E_\gamma$ bin}
Based on the layout of the Hall D beamline,  
the design of the Hall D pair spectrometer, and
the physical location and geometry of the TPOL components provided in Sect.~\ref{sec:construction},
the accuracy of beam polarization measurements performed by TPOL
can be estimated. 
The triplet production beam asymmetry $\Sigma$ discussed in Sect.~\ref{sec:QED}
includes only the effects of triplet photoproduction.
In real materials, recoil electrons passing through
the converter material produce $\delta$-rays,
which also will be seen in the charged particle detector.
Further, $\delta$-rays also will be produced 
by the leptons from pair production in the converter material. 
Rather than the triplet beam asymmetry $\Sigma$, then, 
what is measured in practice by TPOL is the analyzing power $\Sigma_A$,
which includes effects both from triplet production and from $\delta$-rays.
The actual azimuthal variation in yield $Y$ measured experimentally, then, is
\begin{eqnarray}
Y = A[1 - B \cos(2 \phi)]=A[1 - P \Sigma_A \cos(2 \phi)], \label{eqn:yield}
\end{eqnarray}
where $P$ is the photon beam polarization. 
Hence, the photon beam polarization is $P=B/\Sigma_A$,
and $P$ can be derived if the analyzing power $\Sigma_A$ is determined.
This analyzing power will vary based on the photon beam energy $E_\gamma$.

Simulations for evaluating the analyzing power $\Sigma_A$ for TPOL 
for a single energy bin with $E_\gamma=$ 8 to 9 GeV are described here as an example. 
To make such an evaluation,
the components and geometry as constructed and installed
were coded into a GEANT4 simulation of TPOL along with
a realistic intensity profile of the incident photon beam
determined by QED calculations of coherent bremsstrahlung.
The event generator described in Sect.~\ref{sec:QED} was used to model the triplet photoproduction
within the beryllium converter material.
The photon beam strikes the TPOL converter approximately normal to and centered on the converter foil.
The nominal centerline and direction 
of the photon beamline determines the $z$-axis of the simulation coordinate system.
To model production from the photon beam within the converter foil, 
the production vertices for triplet photoproduction events
were distributed uniformly in the $z$-direction throughout the converter thickness, 
while the transverse profile of the vertex distribution used a realistic beam spot
collimated within a diameter of 5 mm.
A 76.2-$\mu$m-thick beryllium converter was simulated.
Generated events were passed through the GEANT4 detector simulation.

In order to model the actual energy resolution of the detector and associated
electronics, the recoil electron energy deposition for each azimuthal sector of the SSD was
randomly reassigned by way of a normal distribution
centered at the observed energy with a standard deviation
of 30 keV (the choice of standard deviation is explained later in
this document).
Within any given simulated TPOL event,
a TPOL hit was defined as a hit where a single azimuthal sector 
was struck by a recoil electron with energy
deposition greater than 160 keV.
Once the number of TPOL hits was established for a simulated event,
only those events with a single TPOL hit having an 
energy deposition greater than 230 keV were further processed.

\begin{figure}[h]
\centering
\includegraphics[width=\linewidth]{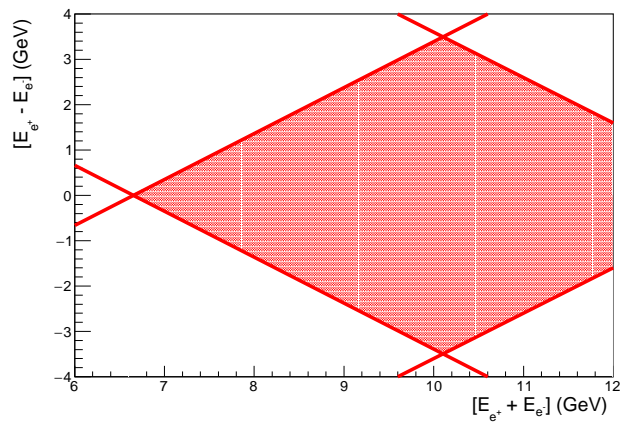}
\caption{{\small{
Plot of the difference $| \Delta E | = | E_+ - E_- |$, 
where $E_+$ is the energy of the pair-production-produced positron and $E_-$ is
the energy of the produced electron, versus  the sum $| E_+ + E_- |$.
The shaded region between
the red lines represents the fiducial region of the 
Hall D pair spectrometer for the Spring 2016 
running of GlueX.
}}}
{\label{fig:deltaEcut}}
\end{figure}

In the simulation, events were required to have an associated pair
energy compatible with the energy acceptance of the Hall D pair spectrometer (PS).
For an event to be considered as valid in this simulation,
the energy difference of the pair $|\Delta(E)|$, 
defined as in Sect.~\ref{sec:design},
had to satisfy the following inequalities:
\begin{eqnarray}
|\Delta(E)| <  \Delta(E_{\rm{max1}})%\nonumber 
\end{eqnarray}
and
\begin{eqnarray}
|\Delta(E)| <  \Delta(E_{\rm{max2}}), %\nonumber
\end{eqnarray}
where
\begin{eqnarray}
\Delta(E_{\rm{max1}}) & \equiv & +1.0 (E_{e^+} + E_{e^-}) - 6.6\ \rm{GeV} \\
\Delta(E_{\rm{max2}}) & \equiv & -1.0 (E_{e^+} + E_{e^-}) + 13.6\ \rm{GeV}. 
\end{eqnarray}
In Hall D of Jefferson Lab, the pair spectrometer was constructed such that 
the accepted $\Delta E$ region as a function of incident
photon energy lay within the red diamond region shown in Fig.~\ref{fig:deltaEcut}. 
(For future running periods of GlueX, the pair spectrometer will be set such that
the maximum acceptance will be near 9 GeV.)

To obtain the azimuthal yield distribution resulting from all these processes, 
the Monte Carlo events passing the criteria described above
had the azimuthal sector hits binned in incident photon energy and azimuthal angle, with
each event weighted by the cross section.
In order to determine the analyzing power $\Sigma_A$, 
the photon beam polarization was set to unity 
(i.e., 100\% linear polarization) solely to minimize the variance of $\Sigma_A$
found from the simulation. 
For each photon energy bin, the azimuthal yield distribution was fit to the function
$A_{t\delta}[1 - B_{t\delta} \cos(2 \phi)]$, where $B_{t\delta}$
represents the beam asymmetry observed for both triplet events
and $\delta$-ray contributions.
To determine the analyzing power $\Sigma_A$, the effects of $\delta$-rays
coming from the leptons resulting from the pair production process also were included
in the simulation.
The pair-induced $\delta$-ray contribution to $\Sigma_A$ was included by treating that effect as a dilution.
Essentially, including $\delta$-rays from the pair production process,
the equation for the yield $Y$ is
\begin{eqnarray}
        Y = A_{t\delta}[1 - B_{t\delta} \cos(2 \phi)] + A_{p\delta}, %\nonumber
\end{eqnarray}
where the yield from the pair-produced $\delta$-rays is given as $A_{p\delta}$.
As noted above in Eq.~\ref{eqn:yield}, this expression also may be written as
\begin{eqnarray}
        Y = A[1 - P \Sigma_A \cos(2 \phi)], %\nonumber
\end{eqnarray}
where, setting $P$ = 1 for these simulations, 
\begin{eqnarray}
        A & \equiv & A_{t\delta} + A_{p\delta}, \nonumber \\
        \Sigma_A & = & B_{t\delta} \frac{A_{t\delta}}{A}. \label{eqn:dilution}
\end{eqnarray}
From the above equations, the analyzing power $\Sigma_A$ is seen to be the
$B_{t\delta}$ asymmetry diluted by a factor of $A_{t\delta}/A$.
Therefore, a dilution factor was defined as $d = A_{t\delta}/A$,
leading to the result that the analyzing power may be written as $\Sigma_A = B_{t\delta} d$.

The value of $A_{p\delta}$ was found by allowing only the created pairs from the
generated triplet events to be processed through the detector simulation. The
integrated detector response, weighted by the triplet cross section and passing
all the analysis cuts, was then scaled by the ratio $R_{pt}=\sigma_p/\sigma_t$,
where $\sigma_p$ ($\sigma_t$) is the pair (triplet) production cross section given by NIST~\cite{NISTxcom}.
For beryllium, $R_{pt}$ is found to vary quite slowly with energy,
from $R_{pt} = 3.38$ at 8 GeV to $R_{pt} = 3.37$  at 10 GeV.

\begin{figure}[t]
\centering
\includegraphics[width=\linewidth]{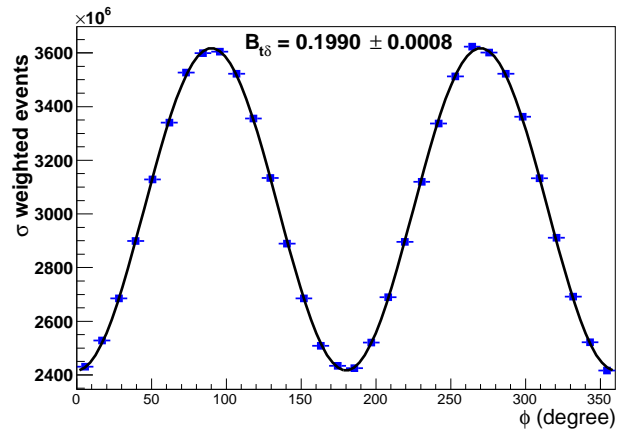}
\caption{{\small{
Analyzing power extraction. The blue points are
cross-section-weighted Monte-Carlo data for triplet events with incident photon beam energies
between 8.5 and 8.6 GeV, including events arising from $\delta-$rays.
The black line represents a fit of the data to $A_{t\delta}[1 + B_{t\delta} \cos(2 \phi)]$,
where $A_{t\delta}$ and $B_{t\delta}$ are parameters of the fit. 
For this energy, the value of $B_{t\delta}$
found has a value of $B_{t\delta} = 0.1990(8)$.
}}}
{\label{fig:anaPower}}
\end{figure}

The yield distribution found from the simulations is shown in Fig.~\ref{fig:anaPower}.
When fit by the expression $Y = A_{t\delta}[1 - B_{t\delta} \cos(2 \phi)] $,
$B_{t\delta}$ was found to be 0.1990 $\pm$ 0.0008. 
The dilution factor determined from simulation was $d = 0.9575$, 
yielding an analyzing power $\Sigma_A = 0.1905(7)$.

In practice, the same procedure is used to determine the analyzing power $\Sigma_A$ 
for a specific photon energy bin $E_\gamma$.
The experimentally measured value of $B(E_\gamma)$ for that photon energy bin will indicate
a photon beam polarization $P$ given by $P(E_\gamma)=B(E_\gamma)/\Sigma_A(E_\gamma)$,
as stated in Eq.~\ref{eqn:yield}.

The primary factors related to the determination of the analyzing power $\Sigma_A$
now have been described. 
While much smaller in impact, an additional refinement 
is added to the analyzing power estimation by including 
the PS geometric acceptance. 
The PS has, in addition to the energy-dependent acceptance discussed above, 
a geometric acceptance for the lepton pairs attributable to
the finite height (3 cm) of the individual PS detector elements. 
This geometric acceptance is also influenced by the nominal distance from the converter
foil to the PS detector elements, taken to be 7.5 m.  
The simulation of these acceptance corrections included a realistic beam spot, 
the nominal distance the created pair travels before striking a PS detector, and the PS detector height. 
When these PS geometric considerations were included, the analyzing power increased by a factor
of 1.0026 for the case where the incident photon was parallel to the floor (PARA); 
for the case where the incident photon was perpendicular to the laboratory floor (PERP), the 
asymmetry increased by a factor of 1.0057. These correction factors must be applied
in the final determination of the beam polarization.

\subsection{Effects from variations in beam spot location}

Systematic uncertainties associated with the 
uncertainty in beam spot location were estimated using simulations.
Within the simulations, the realistic beam spot was
collimated to have a diameter of 5.0 mm and was allowed to be displaced relative to 
the collimator center in steps of 0.1 mm from the nominal position over a grid
with eleven bins in the horizontal ($x_B$) and vertical ($y_B$) directions. In addition
to the beam center displacement relative to the collimator center, the collimator
was varied in vertical displacement relative to the TPOL center. The horizontal (vertical) coordinate of the 
collimator relative to the TPOL center is called $x_0$ ($y_0$). The values of vertical displacement
studied were $y_0$ = 0.0, 0.1, 0.2, 0.3, 0.4, 0.5 mm. In all cases the horizontal displacement of
the collimator relative to the TPOL center was fixed to $x_0$ = 0.0. 

The number of beam spot positions generated relative to the collimator center was 121 and the
number of collimator positions relative to the TPOL center was 6. Thus, the 
total number of beam spot/collimator positions studied was 726. From the set of 726 combinations
of beam spot and collimator positions, the value of $F_0$ was found. 
We take as a conservative estimate of possible 
beam/collimator positions those cases where $F_0 < 0.0225$. The requirement that $F_0 < 0.0225$
reduced the set of 726 combinations of beam spot and collimator positions to 30 cases that
were further analyzed. From the 30 cases of beam/collimator positions compatible with
the restriction that $F_0 < 0.0225$, the largest deviation from the ideal case (100\% polarized
beam with collimator and beam centered on TPOL) was found to be 0.71\%.

\section{Systematic uncertainties in the photon beam polarization estimate} 

Uncertainties in the analyzing power $\Sigma_A$ determined in the previous section
yield a corresponding systematic uncertainty in the beam polarization $P$
made with TPOL. 
The possible sources and estimated contributions of uncertainty in the analyzing power are
described in this section.

\subsection{Converter thickness.}
As discussed above, the passage of the recoil electrons and
the $e^+ e^-$ pair through the converter material results in 
the generation of
$\delta-$rays which strike the SSD. 
The $\delta-$rays are treated as a dilution of the measured beam asymmetry,
as given in Eq.~\ref{eqn:dilution}.
Uncertainties in estimating the dilution factor contributes a 
corresponding systematic uncertainty to the beam asymmetry measurement. 
The number of $\delta-$rays produced and the amount of recoil electron rescattering that occurs
within the converter material both depend upon the thickness of the converter, which 
directly impacts the value of analyzing power.
An uncertainty in the number of $\delta$-rays generated
arises from any uncertainty in the thickness of the converter material. 

The manufacturer of the beryllium converter gives a tolerance of $5\times10^{-4}$ inch 
(12.7 $\rm{\mu m}$) for the stated thickness of the material. 
Assuming this tolerance represents an amount equivalent to three standard
deviations of thickness variation over the photon-illuminated converter, 
then the standard deviation for this thickness measurement 
(and the corresponding variation in the geometry within TPOL) would be 
equivalent to an uncertainty of 4.2 $\rm{\mu m}$. 
To explore the uncertainty resulting from this much variation,
the analyzing power was determined for converter thicknesses of 72.0, 76.2 (nominal) and
80.4 $\rm{\mu m}$. 
The largest percent difference in analyzing power was found from the 
case using the nominal converter thickness and that 
for the 80.4 $\rm{\mu m}$ converter thickness.
Based on this percentage difference,
the estimated systematic uncertainty in the analyzing power $\Sigma$ associated with the
converter thickness was determined to be $\sigma_{\Sigma_A}/\Sigma_A = 0.53\%$.

\subsection{SSD energy threshold and calibration.}
The energy calibration for each sector of the SSD is determined based on
the observed pulse height distribution and the
estimated energy deposited in that sector. 
An energy threshold is required for each particle detected in order for that interaction to 
be classified as a detector hit,
and the value of that energy threshold is based on the energy calibration. 
The energy cut used in the analysis of data and simulation results was 230 keV.
Uncertainties in the energy calibration and the energy resolution of the
device affect the accuracy with which the analyzing power is determined.

The uncertainty in the beam asymmetry coming from the uncertainty in the location of the cut on 
energy deposited in the SSD was determined in the following fashion.
In initial tests, the observed energy deposition spectra  for all sectors of the SSD
were found to be within 80 keV of the Monte Carlo simulations of those spectra,
over an energy deposition range from 0.1 to 3 MeV.
The analyzing power derived with a cut on the energy deposition at 190 keV is 99.1\% of the value
that would have resulted had the cut been placed at 270 keV 
(that is, a difference of 80 keV in the energy cut). 
Therefore, the systematic uncertainty contribution to the analyzing power 
due to uncertainties in the position of the energy cut was taken to be $\sigma_{\Sigma_A}/\Sigma_A = 0.9\%$. 

Based on initial tests, the energy resolution has been observed to have a standard deviation of about 30 keV. 
Taking extreme cases, where the resolution is, at best 10 keV, and at worst 50 keV
the systematic uncertainty associated with energy resolution can be estimated by simulation.
The analyzing power using a 10 keV smear was found to be 0.5\% larger than with a 50 keV
smear. Thus, a conservative estimate for the systematic uncertainty
of the analyzing power due to the energy resolution is taken to be $\sigma_{\Sigma_A}/\Sigma_A = 0.5\%$.

Initial tests found that the sector-by-sector energy calibration was consistent
within a standard deviation of 3 keV. A total of 12 trials were simulated where
the energy cut for a sector came from a random normal distribution of 
standard deviation equal to 3 keV. Using this procedure, 
the uncertainty in the analyzing power due to the sector-by-sector calibration was found to be 0.1\%.

Combining all these SSD energy threshold and energy calibration uncertainties results in an uncertainty of 1.0\%
due to those effects.

\subsection{Detector-to-converter distance.}
The distance from the converter foil to the TPOL detector determines the
minimum and maximum polar angle that the recoil and $\delta-$rays make 
with the detector. Any uncertainty in the distance from the converter foil to
TPOL detector will impact the value of analyzing power.
However, because the SSD is very sensitive, the detector surface cannot
be physically touched by survey equipment once it is positioned in the vacuum chamber.
Instead, the survey determines the distance from the center of the foil to the
downstream face of the PCB that holds the SSD; that distance was found to be 34.9 mm. 
Since the PCB is 2.4 mm thick, the distance from the foil to the center of the PCB is 33.7 mm,
and that estimated distance was used in simulations of an uncertainty in detector-to-converter distance. 
The SSD is 1 mm thick, and its exact placement in the beam direction within the PCB
was taken to be uncertain to $\pm$0.7 mm.
Simulations of the analyzing power for an SSD displaced +0.7 mm and
-0.7 mm from the nominal position in the direction of the incident photon beam
indicated that the analyzing power uncertainty due to this
position uncertainty is 0.54\% . 

\subsection{Photon beam offset.} 
As described above (Sect.~\ref{sc:beamOffset}), any 
transverse offset of the photon beam axis from the nominal center of the TPOL
SSD will affect the value extracted for the analyzing power. Uncertainties in
that position (or offset) thus result in an additional uncertainty in the beam asymmetry extracted.
As discussed in Sect.~\ref{sc:beamOffset} the systematic uncertainty
associated with the beam offset is 0.71\%. 

\subsection{Simulation statistical precision.} 
The finite number of events simulated leads to a statistical
limitation as to how well the analyzing power can be determined from 
the number of simulated events processed during the simulation.
For example, the uncertainty in analyzing power due to simulation statistics over the range
8.4 to 9.0 GeV (the energy range used in \cite{GlueXpaper}) 
was found to be $\sigma_{\Sigma_A}/\Sigma_A = 0.2\%$.
That value can be taken as
the systematic uncertainty in the analyzing power due to the simulation statistical precision.    

\subsection{Beam asymmetry uncertainty.}
The total estimated systematic uncertainty in the analyzing power $\sigma_{\Sigma_A}/\Sigma_A$
based on TPOL arising from all contributions discussed above
is summarized in Table~\ref{tbl:sys}. 
Once each contribution is summed in 
quadrature, the total systematic uncertainty estimate in the beam asymmetry $\sigma_{\Sigma_A}/\Sigma_A$ is 1.5\%.

\begin{table}[h]
\caption{Estimated systematic uncertainty in the 
TPOL determination of beam asymmetry $\Sigma_A$.} \label{tbl:sys}
\centering
\begin{tabular}{cc}
\hline
      Source of     &Estimated \\
       uncertainty  &uncertainty \\
           &($\sigma_{\Sigma_A}/\Sigma_A$) \\
\hline
\hline
Converter thickness & 0.53\%  \\
Threshold, resolution, and calibration & 1.0\% \\
Converter-to-detector distance & 0.54\%  \\
Beam offset & 0.71\% \\
Simulation statistical precision & 0.2\% \\
\\
Total &1.5\%  \\
\hline
\hline
\end{tabular}
\end{table}

\subsection{Beam polarization uncertainty}
In practice, as indicated by Eq.~\ref{eqn:yield},
the photon beam polarization is given by $P = B/\Sigma_A$.
The uncertainty in the photon beam polarization $\sigma_P$ 
is determined by combining the systematic uncertainty for the beam asymmetry $\Sigma_A$ 
described above
(summarized in Table 1 and denoted as $\sigma_{\Sigma_A}$) 
with the statistical uncertainty in the yield ($B$ in Eq.~\ref{eqn:yield}) measured 
during an experiment for the photon energy range of interest.
In the absence of any other sources of uncertainty
(for example, contributions attributable to
radiative corrections to the beam asymmetry for the triplet photoproduction process,
which are expected to be small relative to those uncertainties discussed above), then,
the uncertainty $\sigma_P$ in the photon beam polarization $P$ will be
\begin{eqnarray}
\frac{\sigma_P}{P} = \sqrt{ \left( \frac{\sigma_{\Sigma_A}}{\Sigma_A} \right)^2 +\left( \frac{\sigma_B}{B} \right)^2 }.
\end{eqnarray}

\section{Acknowledgments}
The authors gratefully acknowledge conversations and discussions
with Leonard Maximon and with colleagues in the GlueX Collaboration. 
Work at Arizona State University was supported 
by the National Science Foundation under award PHY-1306737,
and work at the University of Connecticut was supported by
by the National Science Foundation under award PHY-1508238.
Jefferson Science Associates, LLC operated 
Thomas Jefferson National Accelerator Facility for the United States Department of Energy 
under U.S. Department of Energy contract DE-AC05-06OR23177.

\bibliography{NIMrefV2}{}

\end{document}